\documentclass{WileyMSP-template}

\usepackage{lettrine}
\usepackage{amsfonts,amssymb}
\usepackage{amsfonts}
\usepackage{amsmath}
\usepackage{url}
\usepackage{balance}
\usepackage[ruled,linesnumbered]{algorithm2e}
\usepackage{algpseudocode}
\usepackage{amsmath}
\usepackage{graphics}
\usepackage{epsfig}
\usepackage{soul}
\usepackage{color}
\usepackage{float}
\usepackage[utf8]{inputenc}
\usepackage{booktabs}
\usepackage{xcolor}
\usepackage{colortbl}
\usepackage{siunitx}
\usepackage{geometry}
\usepackage{caption}
\usepackage{array}

\definecolor{headergray}{HTML}{EFEFEF}
\definecolor{rowgray}{HTML}{F7F7F7}
\usepackage{verbatim}
\usepackage{cite}
\usepackage[bookmarks=false]{hyperref}
\usepackage{orcidlink}
\usepackage{multirow}
\usepackage{color}
\usepackage{colortbl,hhline}
\usepackage{xcolor}
\usepackage{colortbl}
\usepackage{nomencl}
\makenomenclature
\usepackage{soul}
\usepackage{array}
\usepackage{diagbox}
\usepackage{arydshln}
\usepackage[utf8]{inputenc}
\usepackage[T1]{fontenc}
\usepackage{booktabs}
\usepackage{makecell} 
\usepackage{booktabs} 
\usepackage{multirow}
\usepackage[normalem]{ulem}

\usepackage{geometry}
\usepackage{longtable}
\usepackage{cite}
\usepackage{comment}
\usepackage{url}

\begin{document}

\pagestyle{fancy}

\title{Real-Time Guidewire Tip Tracking Using a Siamese Network for Image-Guided Endovascular Procedures}

\maketitle

\author{Tianliang Yao}
\author{Zhiqiang Pei}
\author{Yong Li}
\author{Yixuan Yuan}
\author{Peng Qi\textsuperscript{*}}

\begin{affiliations}
\textsuperscript{*} Corresponding Author: Peng Qi, email: pqi@tongji.edu.cn)

T. Yao and P. Qi are with the Department of Control Science and Engineering, College of Electronics and Information Engineering, and Shanghai Institute of Intelligent Science and Technology, Tongji University, Shanghai 200092, China;

T. Yao and Y. Yuan are with the Department of Electronic Engineering, Faculty of Engineering, The Chinese University of Hong Kong, Hong Kong SAR 999077, China;


Z. Pei is with the School of Oriental Pan-Vascular Devices Innovation College, University of Shanghai for Science and Technology, 516 Jungong Road, Shanghai 200093, China;

Y. Li is with the Department of Cardiology, the First Affiliated Hospital of Nanjing Medical University, Nanjing 210029, Jiangsu, China;

P. Qi is also with the State Key Laboratory of Cardiovascular Diseases and Medical Innovation Center, Shanghai East Hospital, School of Medicine, Tongji University, Shanghai 200092, China.
\end{affiliations}


\begin{abstract}
\textbf{Abstract:} 
An ever-growing incorporation of AI solutions into clinical practices enhances the efficiency and effectiveness of healthcare services. This paper focuses on guidewire tip tracking tasks during image-guided therapy for cardiovascular diseases, aiding physicians in improving diagnostic and therapeutic quality. A novel tracking framework based on a Siamese network with dual attention mechanisms combines self- and cross-attention strategies for robust guidewire tip tracking. This design handles visual ambiguities, tissue deformations, and imaging artifacts through enhanced spatial-temporal feature learning. Validation occurred on 3 randomly selected clinical digital subtraction angiography (DSA) sequences from a dataset of 15 sequences, covering multiple interventional scenarios. The results indicate a mean localization error of 0.421 ± 0.138 mm, with a maximum error of 1.736 mm, and a mean Intersection over Union (IoU) of 0.782. The framework maintains an average processing speed of 57.2 frames per second, meeting the temporal demands of endovascular imaging. Further validations with robotic platforms for automating diagnostics and therapies in clinical routines yielded tracking errors of 0.708 ± 0.695 mm and 0.148 ± 0.057 mm in two distinct experimental scenarios.
\end{abstract}

\section{Introduction}
\begin{figure}[ht] 
\centering 
\includegraphics[width=\linewidth]{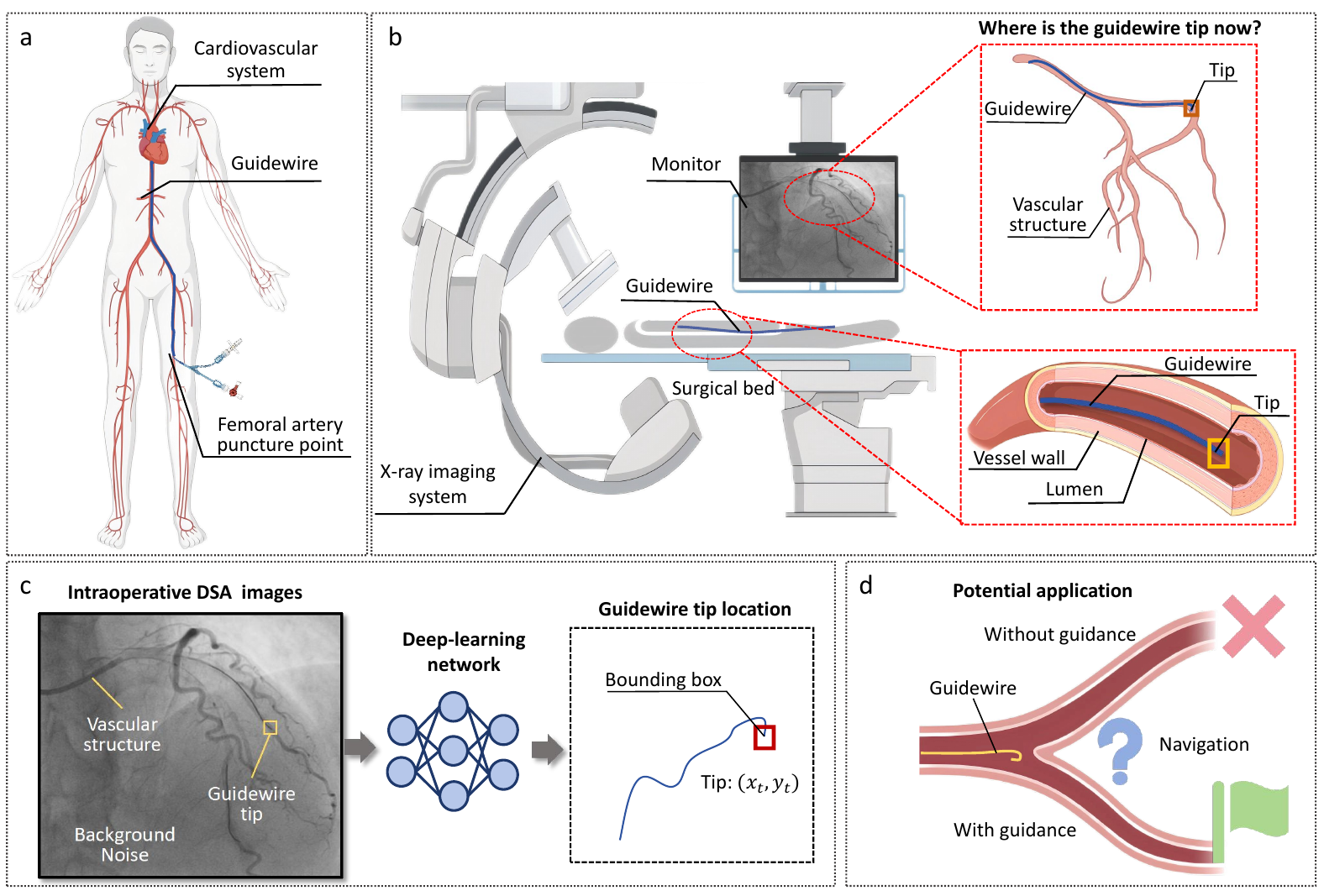} 
\caption{Overview of guidewire tip tracking in endovascular procedures. (a) Anatomical illustration of the cardiovascular system demonstrating the standard guidewire insertion pathway via femoral artery access. (b) Schematic representation of the interventional suite setup, comprising the C-arm X-ray imaging system, surgical bed, and real-time monitoring displays for intraoperative guidance. (c) Representative DSA image annotated with critical elements for automated tracking: guidewire tip position (marked by bounding box), surrounding vascular structures, and anatomical background features that may introduce interference. (d) Comparative illustration of the clinical workflow. The automated tip tracking system provides real-time coordinates $(x_t, y_t)$ to enhance navigation precision through complex vascular networks, potentially reducing procedure time and improving safety. The schematic was created using BioRender (\url{https://biorender.com}).} 
\label{fig:illustration} 
\end{figure}

Throughout the endovascular procedure, the guidewire's role is crucial in ensuring safe and effective access to the target vessel, facilitating precise placement of interventional devices, and minimizing trauma to the vascular system \cite{chen2023catheter}. Its use requires skill and precision to avoid complications such as vessel perforation or dissection. Automatic tip recognition technology for guidewires enhances interventional procedures by enabling real-time localization and improved visualization of the guidewire's tip within the vascular system. This technology optimizes procedural safety by providing real-time feedback, reducing the risk of complications such as vessel perforation or dissection \cite{cruddas2021robotic, yao2025sim2real}. {Fig. \ref{fig:illustration}(a)(b) illustrates the process, including the anatomical pathway, interventional setup, and a representative DSA image. Additionally, it streamlines the procedural workflow, leading to shorter procedure times and reduced radiation exposure. By capturing and analyzing data on guidewire movement and positioning, the technology contributes to procedural quality improvement by enabling precise intraoperative navigation, supports interventional training through accurate movement data capture for simulation environments, and facilitates research in tool-tissue interaction modeling within endovascular systems, ultimately advancing interventional procedures and improving patient outcomes. Meanwhile, Robot-assisted procedures are already being implemented across a diverse range of medical disciplines} \cite{li2024robotic}. Accurately tracking the guidewire tip’s position serves not only to assist physicians in visual navigation but also to provide a reliable perception input for robotic control systems. Since this study further validates the tracking method within a robotic navigation setup, the proposed framework is designed to function as a perception module for autonomous robotic-assisted endovascular procedures. This integration enables downstream motion planning, control feedback, and adaptive tool manipulation in complex anatomical environments.\cite{jingwei2024vascularpilot3d, yao2025sim4endor, scarponi2024autonomous, pore2023autonomous, yao2025advancing}.

Real-time tracking of interventional tools, including needles and guidewires, represents a pivotal research area, with notable advancements across various imaging modalities supporting enhanced precision and safety in minimally invasive procedures \cite{zhang2025mambaxctrack, zhang2025mrtrack}. For instance, studies on photoacoustic imaging have demonstrated the feasibility of high-frame-rate tracking for needles, addressing visibility challenges through deep learning techniques that improve localization accuracy even in noisy environments  \cite{zhao2019minimally, yazdani2021simultaneous, wei2023deep, yan2023visual}. Similarly, ultrasound guidance has shown promise in robot-assisted procedures, achieving improved image quality through the application of visual tracking and motion prediction techniques that enhance the accuracy of needle and guidewire positioning \cite{yan2024task, yan2023learning, yan2021needle}. Furthermore, real-time instrument tracking methods, exemplified by those presented by Ravigopal \textit{et al.} \cite{ravigopal2023real}, have exhibited effective continuum guidewire tracking under fluoroscopic imaging, achieving remarkable robustness by leveraging kinematic modeling to mitigate the impact of imaging noise. 

Despite advancements, existing guidewire tracking methods often depend on accurate segmentation, which poses challenges in clinical settings. DSA, a common real-time imaging modality in endovascular procedures, is prone to artifacts, anatomical variability, and noise \cite{yao2023enhancing, ghibes2025diagnostic}. These disturbances compromise segmentation accuracy, subsequently affecting the robotic system's capacity to provide continuous and precise feedback and adjustments in real-time. Zhou \textit{et al.} \cite{zhou2021real} proposed a multi-task framework utilizing a UNet for guidewire segmentation and endpoint localization, while Chen \textit{et al.} \cite{chen2022automatic} investigated needle detection in 2D ultrasound using deep learning to enhance visibility during insertion. However, segmentation-dependent approaches often struggle with these imaging inconsistencies. For instance, Zhou et al. \cite{zhou2021real} rely on UNet-based segmentation whose performance degrades significantly when vessel boundaries are obscured by motion blur or contrast washout. Similarly, Chen et al. \cite{chen2022automatic} demonstrate detection capabilities that depend heavily on clear needle edge delineation, which may not generalize to low-contrast DSA frames. Consequently, there is a pressing need for a more resilient real-time tracking method that not only addresses these segmentation challenges but also ensures reliable guidewire localization across diverse imaging environments, ultimately enhancing procedural safety and effectiveness in robot-assisted interventions.

To address the critical challenges in guidewire tracking for image-guided endovascular procedures, we propose a novel algorithm based on a Siamese network architecture that integrates spatial attention mechanisms with domain-specific prior knowledge of guidewire tip characteristics in DSA images. Our approach marks the first implementation of a Siamese network specifically tailored for guidewire tip tracking for image-based minimally invasive therapies. By strategically incorporating spatial attention mechanisms, the proposed method dynamically focuses on salient regions of the guidewire tip, effectively mitigating interference from complex background noise, anatomical artifacts, and vascular structures. The algorithm uniquely leverages the distinctive imaging characteristics of guidewire tips in DSA, thereby enabling robust detection and localization capabilities that significantly enhance tracking precision in challenging clinical environments. This innovative approach bridges advanced deep learning techniques with domain-specific medical imaging requirements, presenting a transformative solution for precise guidewire navigation in minimally invasive robotic interventions.

\begin{figure}[htbp]
\centering
\includegraphics[width=0.97\linewidth]{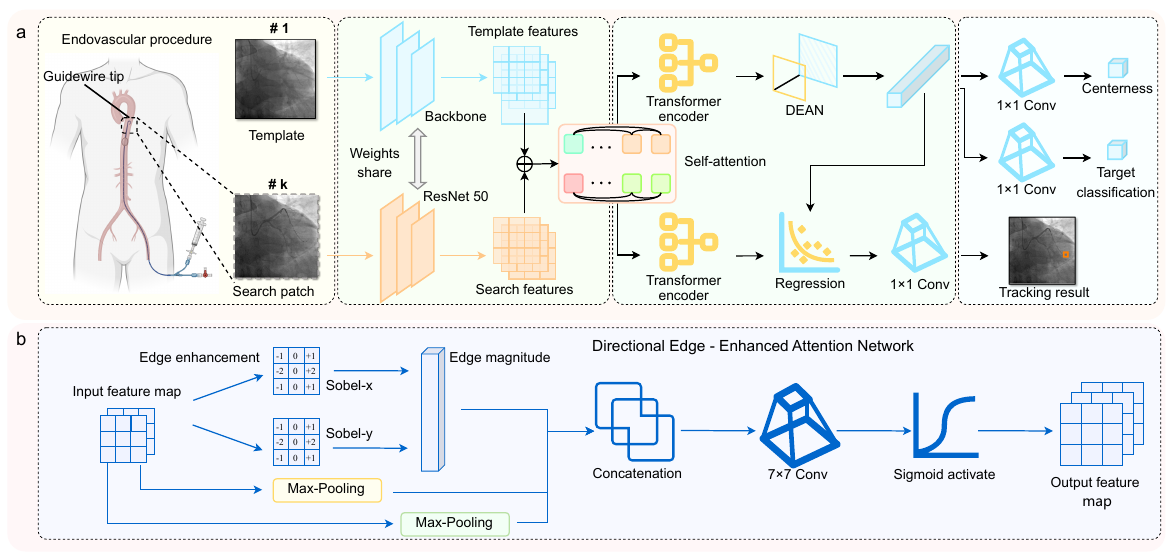}
\caption{Architecture of the proposed Siamese network for guidewire tip tracking in DSA images.  (a) The network consists of a shared-weight ResNet-50 backbone that extracts multi-scale features from both the template image (guidewire tip reference) and search patches. The extracted features are fused via a self-attention mechanism and processed by a transformer encoder. A Directional Edge-Enhanced Attention Network (DEAN) further refines the features to emphasize guidewire-specific geometric patterns.The output is fed into multiple heads to generate target classification, centerness scores, and regression-based localization results.(b) The DEAN module enhances edge features by computing directional edge maps using Sobel filters and max-pooling, which are concatenated with input features and passed through a convolutional layer followed by a sigmoid activation to produce the output.}
\label{fig:Framework}
\end{figure}

\section{Methodology}
\subsection{Problem Formulation}
Given a DSA image sequence $\{\mathcal{I}_t\}_{t=1}^T$, guidewire tip tracking aims to estimate the tip location \\ $\mathbf{b}_t = [x_t, y_t, w_t, h_t] \in \mathbb{R}^4$ in each frame, representing the bounding box coordinates and dimensions. Let $\mathcal{Z}_t \in \mathbb{R}^{H \times W}$ denote the template patch centered at the previous prediction $\mathbf{b}_{t-1}$, and $\mathcal{X}_t \in \mathbb{R}^{H \times W}$ represent the search region in the current frame. The objective is to learn a mapping function $\mathcal{F}: (\mathcal{Z}_t, \mathcal{X}_t) \to (\mathbf{b}_t, r_t)$, where $\mathbf{b}_t$ specifies the tip location and $r_t \in [0,1]$ indicates the prediction confidence.

The framework is designed to address the inherent challenges of DSA imaging, including low contrast between the guidewire and background, motion blur due to rapid guidewire movement or patient motion, complex vascular overlaps, and noise artifacts such as those caused by anatomical variability or imaging equipment \cite{yao2023enhancing}. These challenges often lead to visual ambiguities, where the guidewire tip may be obscured by overlapping vascular structures or diminished by low-contrast conditions, complicating accurate localization. To mitigate these issues, the proposed Siamese network-based approach employs a template-matching paradigm that leverages the temporal consistency of the guidewire’s appearance across frames. By using the template patch \mbox{$\mathcal{Z}_t$} as a reference, the model establishes a robust correlation with the search region \mbox{$\mathcal{X}_t$}, enabling precise localization even in the presence of dynamic deformations and background interference~\cite{cheng2021learning}.

The mapping function is defined as:

\begin{equation}
\mathbf{b}_t, r_t = \mathcal{F}(\mathcal{Z}_t, \mathcal{X}_t; \theta),
\end{equation}

where $\theta$ denotes the parameters of the mapping function. The template $\mathcal{Z}_t$ and search region $\mathcal{X}_t$ are processed to produce a feature representation, enabling precise localization despite visual ambiguities. This approach facilitates reliable guidewire tip tracking in dynamic and low-contrast environments.

\subsection{Network Architecture}
As illustrated in Fig.~\ref{fig:Framework}, the proposed method employs a Siamese architecture with shared parameters to process template and search images, ensuring consistent feature extraction~\cite{li2022survey}. The architecture integrates a modified ResNet-50 backbone, a multi-scale feature extraction mechanism, a Transformer encoder, and a Directional Edge-Enhanced Attention Network (DEAN), each contributing to fine-grained feature extraction and long-range dependency modeling to address the challenges of low-contrast and complex vascular environments.

The feature extraction process is defined as:
\begin{equation}
\begin{aligned}
\mathbf{F}_z &= \phi(\mathcal{Z}_t; \theta) \in \mathbb{R}^{C \times H_z \times W_z}, \\
\mathbf{F}_x &= \phi(\mathcal{X}_t; \theta) \in \mathbb{R}^{C \times H_x \times W_x},
\end{aligned}
\end{equation}
where $\phi(\cdot; \theta)$ represents the feature extraction function parameterized by $\theta$, and $\mathcal{Z}_t$ and $\mathcal{X}_t$ denote the template and search images, respectively.

Multi-scale feature extraction is achieved by leveraging convolutional outputs from ResNet-50 at $L=4$ levels, enabling the capture of fine-grained details and broader contextual information:
\begin{equation}
\{\mathbf{F}^l\}_{l=1}^L = \mathcal{F}_{\text{backbone}}(\mathcal{I}; \theta), \quad \mathbf{F}^l \in \mathbb{R}^{C_l \times H_l \times W_l},
\end{equation}
where $C_l$, $H_l$, and $W_l$ denote the channel count, height, and width of the feature map at level $l$. This multi-scale approach is critical for fine-grained feature extraction, as it captures the subtle geometric details of the guidewire tip (e.g., its thin, elongated structure) at higher-resolution layers while incorporating contextual vascular patterns at lower-resolution layers. This ensures robust localization despite low contrast and background interference in DSA images.

The Transformer encoder processes the fused features from the template and search images to model long-range dependencies, capturing spatial relationships across complex vascular structures. By applying self-attention mechanisms, the encoder enhances the network’s ability to focus on relevant regions, such as the guidewire tip, while modeling interactions between distant anatomical features. This is particularly effective in handling vascular overlaps and noise artifacts, as it allows the network to contextualize the guidewire’s position relative to the broader vascular environment. The Transformer encoder’s output is then fed into the DEAN module for further refinement.

The DEAN module, detailed in Section 2.3, enhances guidewire-specific geometric patterns by integrating directional edge filtering with dual attention mechanisms. It emphasizes the guidewire’s linear and elongated structures, suppressing irrelevant background textures and artifacts. This targeted enhancement complements the multi-scale feature extraction by refining fine-grained details and supports the Transformer encoder’s long-range dependency modeling by focusing attention on guidewire-relevant regions. Together, these components enable precise and robust guidewire tip localization in challenging DSA imaging conditions.

\subsection{Directional Edge-Enhanced Attention Network}

The Directional Edge-Enhanced Attention Network (DEAN) is designed to selectively amplify guidewire-relevant features while suppressing background interference in DSA images. To achieve this, the module integrates directional edge filtering with a dual attention mechanism, enabling fine-grained enhancement of linear and spatially constrained structures such as guidewires.

The edge enhancement stream employs Sobel filters to detect guidewire-like linear structures based on local directional gradients, chosen for their computational efficiency and robustness in the context of DSA imaging. The input feature map $\mathbf{F}$ is processed through a bank of learnable-oriented filters, defined as:
\begin{equation}
\mathbf{E} = \mathcal{G}(\mathbf{F}; \Phi) = \{\mathbf{F} * \mathbf{g}_k\}_{k=1}^{K},
\end{equation}
where $\{\mathbf{g}_k\}_{k=1}^{K}$ denotes a set of $K$ directional convolution kernels with learnable parameters $\Phi$, and $*$ represents the convolution operation. Sobel filters are selected over traditional methods like Canny Edge Detection and HED-based approaches due to their specific advantages for guidewire tracking. Unlike Canny Edge Detection, which involves multi-stage processing (e.g., noise reduction, gradient computation, non-maximum suppression, and hysteresis thresholding), Sobel filters provide a simpler, single-pass gradient computation that is less sensitive to noise in low-contrast DSA images, reducing the risk of fragmented or false edges around the guidewire. Compared to HED-based approaches, which rely on deep learning to capture complex edge patterns and require extensive training data and computational resources, Sobel filters offer a lightweight solution that aligns with the real-time processing demands of endovascular procedures. The learnable nature of the directional kernels in DEAN further enhances their adaptability, allowing optimization for the guidewire’s elongated geometry while attenuating irrelevant textures and noise.

To further refine the representation, a dual attention mechanism is applied. It consists of a channel attention module and a spatial attention module, both of which operate on the original feature map $\mathbf{F}$ and the edge-enhanced map $\mathbf{E}$. The channel attention weight $\mathbf{A}_c$ is computed by aggregating global semantic context using global average pooling (GAP), followed by a fully connected transformation:
\begin{equation}
\mathbf{A}_c = \sigma(\mathcal{W}_c[\text{GAP}(\mathbf{F}) \| \text{GAP}(\mathbf{E})]),
\end{equation}
where $\sigma(\cdot)$ denotes the sigmoid activation function, $\mathcal{W}_c$ is a trainable projection layer, and $\|$ represents feature concatenation. This mechanism emphasizes the most informative channels relevant to guidewire detection.

Simultaneously, spatial attention is computed by combining local appearance cues from both $\mathbf{F}$ and $\mathbf{E}$:
\begin{equation}
\mathbf{A}_s = \sigma(\mathcal{W}_s[\mathbf{F} \| \mathbf{E}]),
\end{equation}
where $\mathcal{W}_s$ is a convolutional layer that encodes spatial correlations. The resulting attention map $\mathbf{A}_s$ highlights regions that are likely to contain the guidewire tip or shaft.

The final enhanced feature representation $\mathbf{F}_{\text{out}}$ is obtained through multiplicative fusion of the input feature map with both attention maps:
\begin{equation}
\mathbf{F}_{\text{out}} = \mathbf{F} \odot \text{Broadcast}(\mathbf{A}_c) \odot \mathbf{A}_s,
\end{equation}
where $\odot$ denotes element-wise multiplication and $\text{Broadcast}(\cdot)$ expands the channel-wise attention vector to match the spatial dimensions of $\mathbf{F}$.

This architecture enables the network to selectively attend to elongated, low-contrast anatomical structures under noisy imaging conditions. The combined edge-aware and attention-guided mechanisms allow for improved localization of the guidewire tip and better discrimination from surrounding vascular structures and artifacts.

\subsection{Cross-Correlation and Prediction}

To efficiently measure the similarity between the guidewire template and candidate regions in the search image, a frequency-domain cross-correlation operation is employed. This approach enables fast computation while preserving translation equivariance, which is essential for accurate localization under shifting field-of-view conditions. The correlation response map $\mathbf{R}$ is computed as:
\begin{equation}
\mathbf{R} = \mathcal{F}^{-1}\left(\frac{\hat{\mathbf{F}}_z^* \odot \hat{\mathbf{F}}_x}{|\hat{\mathbf{F}}_z|^2 + \epsilon}\right),
\end{equation}
where $\mathcal{F}^{-1}(\cdot)$ denotes the inverse Fourier transform, $\hat{\mathbf{F}}_z$ and $\hat{\mathbf{F}}_x$ represent the Fourier-transformed feature maps of the template and search region, respectively, $(\cdot)^*$ indicates complex conjugation, $\odot$ denotes element-wise complex multiplication, and $\epsilon$ is a small regularization constant to avoid division by zero.

The output response map $\mathbf{R}$ provides a dense similarity score for each spatial location within the search region, with high values indicating strong correspondence to the template. This correlation map serves as the input to the subsequent prediction heads, which jointly estimate target presence, confidence, and location refinement.

Specifically, the prediction module consists of three branches: a classification head for presence detection, a centerness head to assess the spatial confidence of localization, and a regression head to refine the coordinates of the predicted bounding box. Let $\hat{r}_t$, $\hat{c}_t$, and $\hat{\mathbf{b}}_t$ denote the outputs of the three branches, respectively, corresponding to classification, centerness, and bounding box prediction at time $t$. These outputs are supervised using the multi-task loss function provided below.

\subsection{Multi-Task Learning Objective}

The proposed network is trained under a unified multi-task learning framework designed to concurrently optimize spatial localization accuracy, classification confidence, and regression consistency. The choice of loss functions—Generalized Intersection over Union (GIoU), Binary Cross-Entropy (BCE), and Smooth L1—is tailored to address the dynamic challenges of guidewire tip tracking in DSA sequences, including rapid motion, sharp direction changes, and the need for stable IoU and precise boundary alignment. The total loss function is defined as:

\begin{equation} 
\mathcal{L} = \lambda_1 \mathcal{L}_{\text{loc}} + \lambda_2 \mathcal{L}_{\text{cls}} + \lambda_3 \mathcal{L}_{\text{reg}}, 
\end{equation}

where $\lambda_1$, $\lambda_2$, and $\lambda_3$ denote the weighting coefficients that control the relative influence of each loss term.

The localization loss $\mathcal{L}_{\text{loc}}$ is computed using the Generalized Intersection over Union (GIoU) metric, which extends the standard IoU by incorporating the shape and position relationship between predicted and ground truth boxes. This formulation improves localization performance in cases of partial or non-overlapping predictions:
\begin{equation} 
\mathcal{L}_{\text{loc}} = 1 - \text{GIoU}(\mathbf{b}_t, \hat{\mathbf{b}}_t), 
\end{equation}
where $\mathbf{b}_t$ and $\hat{\mathbf{b}}_t$ represent the ground truth and predicted bounding box coordinates at time $t$, respectively. GIoU is particularly suited for guidewire tracking under rapid motion, as it accounts for the smallest convex hull enclosing both boxes, ensuring stable IoU even when the guidewire tip moves quickly or undergoes sharp direction changes. This robustness to non-overlapping scenarios enhances boundary alignment, critical for maintaining tracking precision in dynamic DSA environments.

The classification loss $\mathcal{L}_{\text{cls}}$ is formulated using the binary cross-entropy function between the predicted classification confidence score $\hat{r}_t$ and the binary ground truth label $r_t$:
\begin{equation} 
\mathcal{L}_{\text{cls}} = -\left[r_t \log(\hat{r}_t) + (1 - r_t) \log(1 - \hat{r}_t)\right]. 
\end{equation}
BCE is chosen for its effectiveness in optimizing classification confidence, ensuring reliable detection of the guidewire tip amidst background noise and vascular overlaps. Its stability in dynamic scenarios, where rapid motion may introduce visual ambiguities, supports consistent identification of the guidewire tip, contributing to overall tracking robustness.

The regression loss $\mathcal{L}_{\text{reg}}$ employs the Smooth L1 loss to enhance numerical stability and mitigate sensitivity to outliers in bounding box coordinate regression:
\begin{equation}
\mathcal{L}_{\text{reg}} = \text{smooth}_{\text{L1}}(\mathbf{b}_t - \hat{\mathbf{b}_t)},
\end{equation}

where the Smooth L1 function is applied element-wise to each bounding box coordinate. This loss function is selected for its balanced sensitivity to both small and large errors, making it well-suited for precise boundary regression in the presence of sharp direction changes. Unlike L2 loss, which can overpenalize large deviations, Smooth L1 provides a smooth gradient for small errors, enabling fine-grained adjustments to the bounding box coordinates, which is essential for tracking the guidewire tip under rapid and unpredictable motion. This ensures accurate boundary alignment even in challenging deformation scenarios.

The combination of GIoU, BCE, and Smooth L1 losses is particularly effective for guidewire tracking, as it addresses the unique demands of dynamic DSA imaging. GIoU ensures stable localization, BCE maintains reliable classification, and Smooth L1 supports precise regression, collectively handling the challenges of motion and deformation. The loss functions are tailored through empirical weight tuning of \mbox{$\lambda_1, \lambda_2, \lambda_3$}, determined via grid search, to optimize performance across diverse DSA datasets, ensuring robustness to varying motion patterns and clinical conditions. 

The loss weights $\lambda_1$, $\lambda_2$, and $\lambda_3$ are selected empirically through grid search to ensure convergence across diverse anatomical geometries and imaging conditions. This multi-objective formulation enables the network to simultaneously learn precise localization, reliable detection confidence, and consistent spatial regression, which are critical for real-time clinical deployment.

\begin{figure}[ht]
\centering
\includegraphics[width=0.84\linewidth]{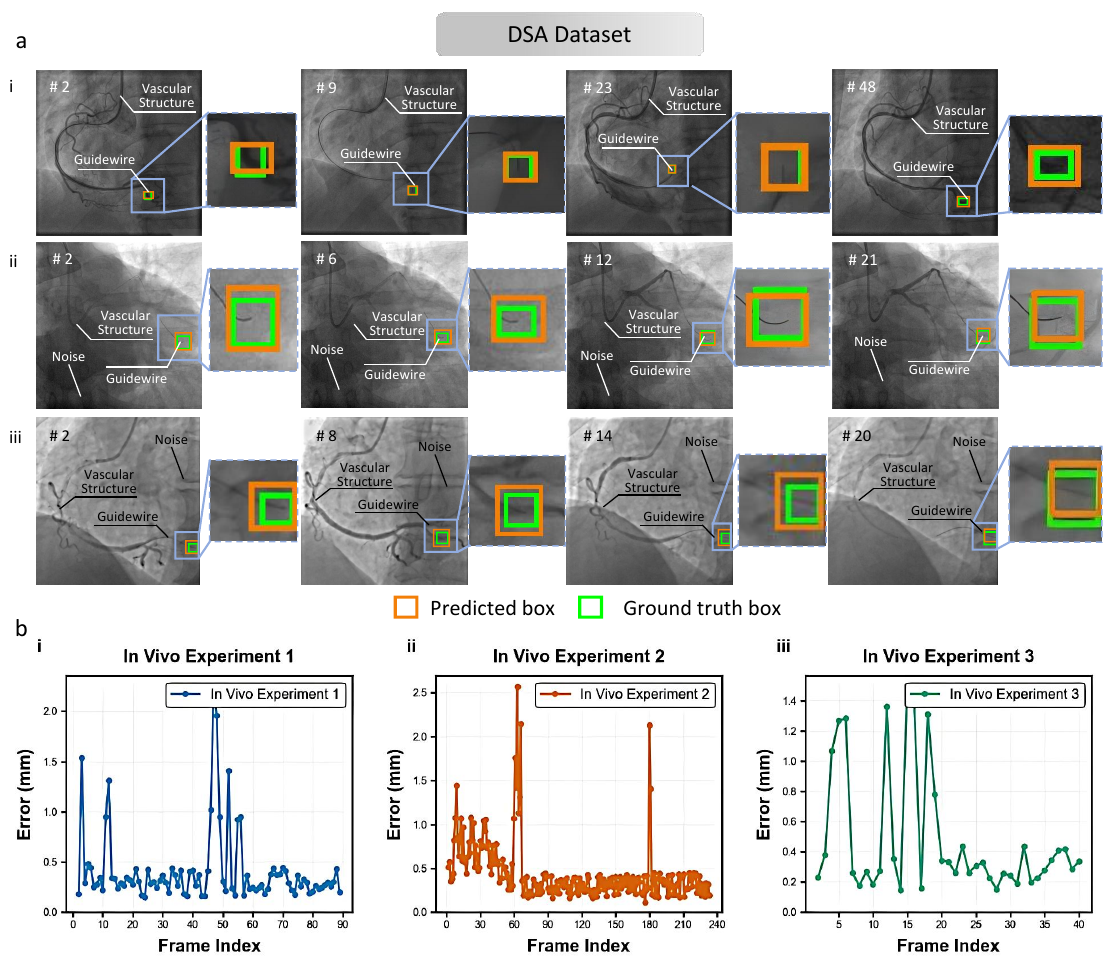}
\caption{Qualitative and quantitative results of our method on in-vivo experiments of DSA dataset. (a) Visualization results of guidewire tip tracking. The orange boxes represent the predicted bounding boxes while the green boxes show the ground truth. Background noise such as bone structures and other anatomical features are present in the images. (i) Results from in-vivo experiment 1, showing robust detection despite various vascular structures and background noise. (ii) Results from in-vivo experiment 2, demonstrating detection performance under different anatomical interference. (iii) Results from in-vivo experiment 3, illustrating detection capability with complex background structures. (b) IoU error curves comparing the predicted boxes with ground truth over frame sequences. (i) Error curve for in-vivo experiment 1. (ii) Error curve for in-vivo experiment 2. (iii) Error curve for in-vivo experiment 3.}
\label{fig:DSA_Tracking_Images}
\end{figure}

\section{Performance Validation}
\subsection{Experimental Setup}
The dataset consists of 15 clinical DSA sequences from real endovascular procedures across 15 patients. It totals 1347 frames. These sequences originate from coronary interventions. They target the left anterior descending, right coronary, and circumflex arteries. The data captures patient-specific anatomical variations. The frames exhibit variability from guidewire motion and contrast agent flow. This affects vessel visibility and background intensity. The dataset is split into 12 training sequences with 1078 frames and 3 testing sequences with 269 frames. No overlap exists between them. Testing sequences are randomly selected for diversity in vascular structures and scenarios. Ground truth annotations are created by two cardiologists using LabelMe. They mark guidewire tips with bounding boxes averaging 12 × 12 pixels with a standard deviation of ±3 pixels. A double-confirmation protocol ensures accuracy through cross-validation.

The robotic integration experiments utilize a custom endovascular intervention robot from the United Imaging Research Institute. It features precise guidewire manipulation via a leader-follower control system. The setup includes a validated vascular phantom with radio-opaque markers. This phantom replicates coronary artery anatomy, including vessel diameters, curvatures, and bifurcations, based on clinical angiographic data. Ground truth positions are provided by the robot's registration system. This system achieves ±0.1 mm accuracy. An RGB camera uses strong exposure and high contrast. Grayscale preprocessing preserves DSA-like features. The DSA-trained model applies zero-shot to this setup. It tests adaptability. Two navigation tasks assess real-time tracking and robustness. They evaluate the potential for enhancing robotic-assisted endovascular interventions.

\subsection{Implementation Details}
Our tracking framework employs a Siamese architecture based on a modified ResNet-50 backbone, pre-trained on ImageNet and specifically adapted for DSA imaging characteristics. The network processes paired template-search frames for temporal correlation learning. To enhance feature representation in DSA-specific scenarios, we implemented several architectural modifications: (1) Custom adjustment layers at multiple scales to handle the unique challenges of guidewire tracking in vascular structures, (2) A specialized classification head utilizing Composite Anchors Regression for precise localization, and (3) Group Normalization ($32$ groups) with Leaky ReLU activations to maintain stable training in regions with varying contrast and intensity gradients.

The training process employed SGD optimization with momentum ($\mu = 0.9$) and weight decay ($\lambda = 1 \times 10^{-4}$), complemented by a cosine annealing learning rate schedule. To enhance model robustness and generalization, we implemented domain-specific data augmentation strategies, including random rotations ($\pm30^{\circ}$), scaling ($0.9$-$1.1$), and intensity variations ($\pm10\%$), specifically designed to simulate realistic variations in clinical DSA imaging conditions.

\begin{table}[!t]
\renewcommand{\arraystretch}{1.2}
\caption{Performance of Tracking Methods Across In Vivo Datasets}
\label{tab:tracking_performance}
\centering
\footnotesize
\begin{tabular}{@{}l *{5}{S[table-format=1.3]} c c S[table-format=1.3]@{}}
\toprule
\rowcolor{headergray}
\textbf{Method} & {\textbf{Mean Err.}} & {\textbf{Std. Dev.}} & {\textbf{Min Err.}} & {\textbf{Max Err.}} & {\textbf{FPS}} & {\textbf{Filter Type}} & {\textbf{DEAN Aug.}} & {\textbf{IoU}} \\
\rowcolor{headergray}
& {(\si{\mm})} & {(\si{\mm})} & {(\si{\mm})} & {(\si{\mm})} & & & & \\
\midrule
Proposed (Avg.) & 0.421 & 0.138 & 0.116 & 1.736 & 57.2 & None & Yes & 0.782 \\
\rowcolor{rowgray}
Baseline (Avg.) & 0.784 & 0.231 & 0.137 & 1.512 & 45.6 & None & No & 0.641 \\
Kalman \cite{yan2021needle} & 0.652 & 0.194 & 0.134 & 1.241 & 62.5 & Linear & No & {-} \\
\rowcolor{rowgray}
Ext. Kalman \cite{tong2023uniting} & 0.573 & 0.167 & 0.102 & 1.053 & 58.3 & Nonlinear & No & {-} \\
Particle \cite{iovene2024hybrid} & 0.491 & 0.152 & 0.134 & 1.894 & 55.7 & Nonparam. & No & {-} \\
\rowcolor{rowgray}
Unsc. Kalman \cite{wang2025active} & 0.534 & 0.161 & 0.103 & 1.987 & 56.9 & Nonlinear & No & {-} \\
\midrule
\multicolumn{9}{@{}l}{\footnotesize \textit{Note:} Results averaged across three in vivo datasets.} \\
\bottomrule
\end{tabular}
\end{table}

\begin{figure}[ht]
\centering
\includegraphics[width=\linewidth]{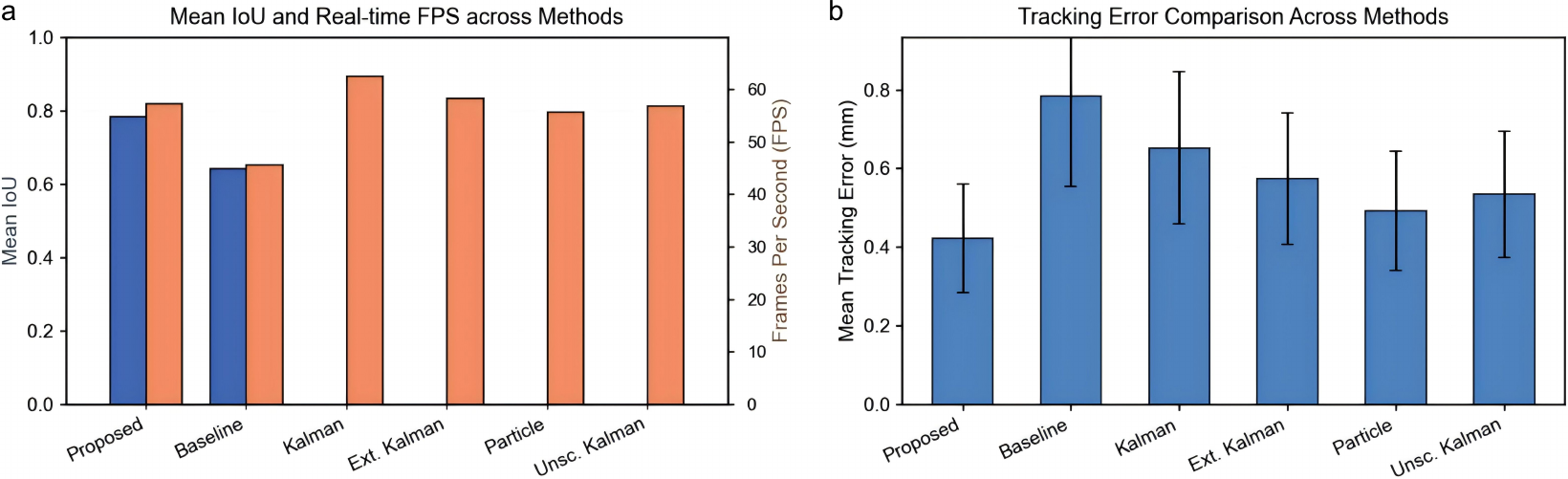}
\caption{Comprehensive comparison of tracking performance and computational efficiency across methods. (a) Quantitative comparison of mean Intersection over Union (IoU, blue bars) and real-time processing speed (FPS, orange bars) for each evaluated method. IoU values are only shown for methods where this metric is applicable. (b) Mean tracking error (mm) with standard deviation (error bars) for each method, evaluated on in vivo datasets.}
\label{fig:Visulization}
\end{figure}

\begin{table}[ht]
\renewcommand{\arraystretch}{1.2}
\caption{Tracking performance metrics for robotic navigation experiments.}
\label{tab:robotic_tasks}
\centering
\footnotesize
\begin{tabular}{@{}lccc p{10.0cm}@{}}
\toprule
\textbf{Task} & \textbf{Mean Error (mm)} & \textbf{Std. Dev. (mm)} & \textbf{FPS} & \textbf{Description} \\
\midrule
Task 1 & 0.708 & 0.695 & 38 & Multi-branch navigation with frequent bifurcations and direction changes. \\
Task 2 & 0.148 & 0.057 & 36 & Retraction path with fewer bifurcations. \\
\bottomrule
\end{tabular}
\end{table}

\subsection{Evaluation Protocol}
Performance evaluation was conducted on the selected clinical DSA sequences, focusing on challenging scenarios including vessel overlapping, varying contrast conditions, and dynamic motion patterns. The quantitative assessment utilized two primary metrics: (1) Intersection over Union (IoU) for measuring tracking precision, and (2) Center error distance in mm, both calculated against the expert-annotated ground truth. The output bounding box is defined as $\mathbf{b}_t = [x_t, y_t, w_t, h_t]$, where $(x_t, y_t)$ denotes the center, and $w_t$ and $h_t$ represent width and height. Center error measures the Euclidean distance between the predicted center $(x_t, y_t)$ and the ground truth center, assuming both positions the guidewire tip at the box center. IoU assesses overlap based on the full box dimensions. To evaluate real-time performance capabilities, we measured the processing speed in frames per second (FPS) on our experimental platform, as interventional procedures typically require a minimum of 15-30 FPS for smooth visualization and effective guidance \cite{hatt2015robust}.

In the robotic integration setting, we further assessed the temporal consistency of tracking in 3D space and system latency for real-time performance evaluation. The zero-shot transfer performance was evaluated by evaluating tracking accuracy between clinical and robotic settings without additional training or fine-tuning, demonstrating the model's generalization capabilities across different deployment scenarios.

\section{Results}
\subsection{Precision and Robustness of Guidewire Tracking}

The performance of the proposed guidewire tracking method is evaluated against established filtering techniques, with results summarized in Table~\ref{tab:tracking_performance}, Table~\ref{tab:robotic_tasks} and Fig. \ref{fig:Visulization}. The proposed method achieves a mean tracking error of 0.421 mm with a standard deviation of 0.138 mm, surpassing conventional approaches. Specifically, it outperforms the Kalman Filter (0.652 mm ± 0.194 mm), Extended Kalman Filter (0.573 mm ± 0.167 mm), Particle Filter (0.491 mm ± 0.152 mm), and Unscented Kalman Filter (0.534 mm ± 0.161 mm), yielding a 14\% to 35\% reduction in mean error relative to these baselines. The low standard deviation of the proposed method indicates consistent tracking performance across frames, a critical requirement for clinical applications.

Linear and nonlinear filtering methods, such as the Kalman and Extended Kalman Filters, exhibit higher errors and variability in complex medical imaging scenarios, such as dynamic DSA. While the Particle Filter and Unscented Kalman Filter demonstrate improved performance over traditional Kalman-based methods, they are still outperformed by the proposed adaptive approach. Integrating the DEAN framework, the proposed method achieves a mean Intersection over Union (IoU) of 0.782 while maintaining a processing speed of 57.2 frames per second (FPS). In contrast, the baseline average yields a mean error of 0.784 mm ± 0.231 mm at 45.6 FPS with an IoU of 0.641. These results highlight the proposed method’s superior precision, robustness, and computational efficiency, making it well-suited for real-time interventional procedures.
\subsection{Visualization and Detection Capabilities}

Fig.\ref{fig:DSA_Tracking_Images} presents a detailed evaluation of the guidewire tip detection and tracking performance based on both qualitative visual results and quantitative error analysis. Fig.\ref{fig:DSA_Tracking_Images}(a) illustrates selected frames from a representative DSA dataset, encompassing diverse anatomical regions and imaging conditions. Each row (i–iii) corresponds to distinct experimental sequences, where guidewire localization is challenged by overlapping vascular structures, radiopaque noise artifacts, and low contrast boundaries. The guidewire tip regions are magnified for clarity, and bounding boxes indicate the predicted positions (orange) and ground truth annotations (green). Across all scenarios, the predicted positions exhibit high spatial consistency with the annotated ground truth, demonstrating the method's resilience to contrast variability and vascular occlusion.

Specifically, in subfigure~(a-i), the guidewire tip is consistently identified within frames exhibiting strong vascular contrast, with precise overlap of predicted and true locations. In subfigure~(a-ii), performance remains stable despite interference from noise signals and reduced background contrast. Subfigure~(a-iii) highlights frames from sequences with more complex backgrounds, where the system maintains accurate detection despite diminished guidewire visibility.

Fig. \ref {fig:DSA_Tracking_Images}(b) provides the frame-wise localization error in millimeters for three separate in vivo experiments. The plots (i–iii) reflect temporal error fluctuations, which remain within a narrow range for the majority of frames. In particular, Fig.~\ref {fig:DSA_Tracking_Images}~(b-i) shows a consistently low error margin across 90 frames, indicating high tracking fidelity. In Fig.~\ref {fig:DSA_Tracking_Images}~(b-ii), although the initial frames exhibit higher error due to sudden motion or image noise, the method rapidly stabilizes. Fig. \ref {fig:DSA_Tracking_Images}~(b-iii) demonstrates reliable performance under conditions involving intermittent occlusion and background noise.

\begin{figure}[htbp]
\centering
\includegraphics[width=0.93\linewidth]{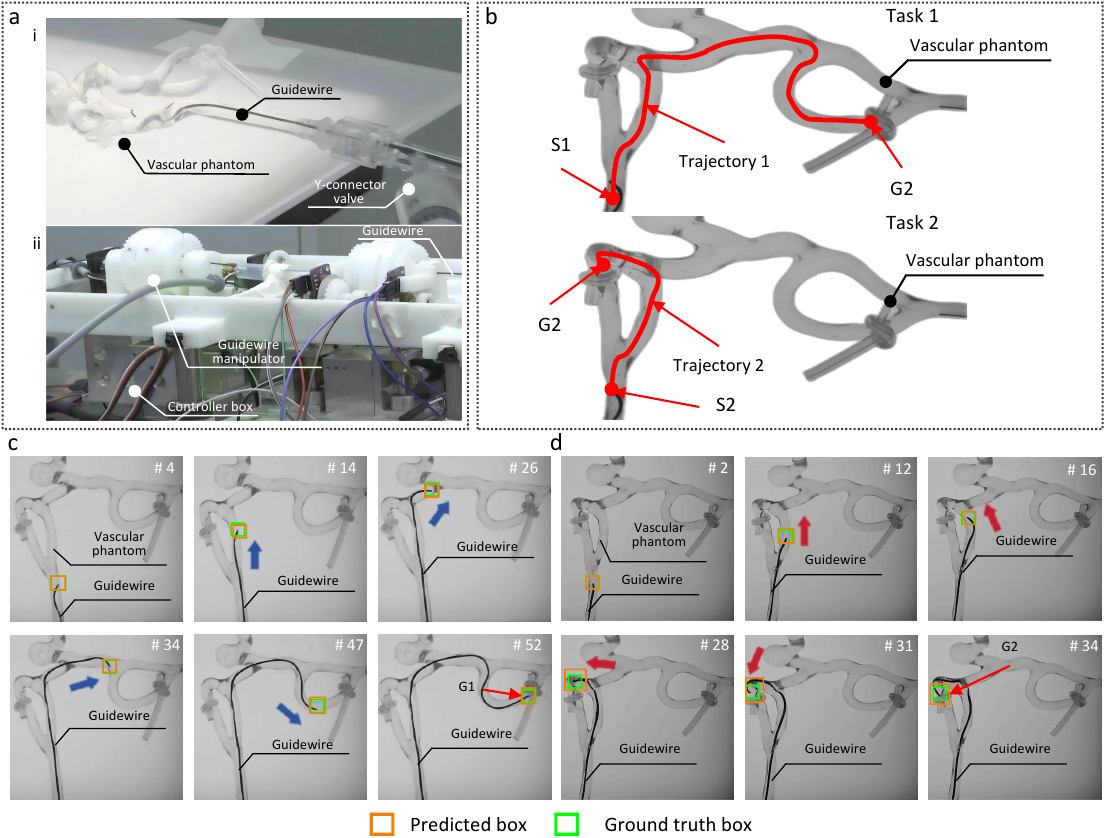}
\caption{Overview of the robotic guidewire navigation system and experimental validation results. (a) System configuration showing the controller box, guidewire manipulator, and vascular phantom with Y-connector valve for automated guidewire navigation. (b) Two predefined robotic guidewire navigation tasks are illustrated in the vascular phantom model. (c) Task 1: Autonomous navigation from start point S1 to goal point G1 through multiple vessel bifurcations. (d) Task 2: Autonomous navigation from start point S2 to goal point G2 with fewer bifurcations but requiring guidewire retraction maneuvers. The orange boxes represent manually labeled ground truth positions, while green boxes indicate algorithm-predicted guidewire tip positions.}
\label{fig:Deployment}
\end{figure}

\subsection{Computational Efficiency and Advanced Augmentation}
The proposed method integrates computational efficiency with machine-learning techniques to support reliable tracking in interventional medical procedures. Operating at 57.2 frames per second, it meets the real-time processing demands of such applications while ensuring consistent performance. The use of DEAN augmentation enhances the system’s adaptability to varied imaging conditions, improving generalization.

Compared to the baseline method, which recorded an IoU of 0.641, the proposed approach achieves a mean IoU of 0.782. This improvement reflects the effectiveness of combining adaptive filtering with data augmentation to address challenges in guidewire dynamic tracking. The method provides a practical solution for accurate localization in complex medical environments, supporting autonomous navigation in image-guided endovascular interventions.

\subsection{Integration with Robotic System}

The proposed vision-based guidewire tip tracking algorithm was integrated into a robotic navigation platform for experimental validation under controlled vascular phantom conditions. The system architecture is depicted in Fig.~\ref{fig:Deployment}(a), which includes a modular controller box, an automated guidewire manipulator, and a 3D-printed vascular phantom incorporating a Y-connector valve for procedural realism. The guidewire manipulator is actuated by servo motors, enabling advancement, retraction, and axial rotation, with motion commands governed by a leader–follower control scheme. During navigation, the robotic system receives guidewire tip coordinates in real time from the tracking module, which extracts the centroid of the predicted bounding box for continuous trajectory adjustment within the vascular environment \cite{yao2025sim2real, scarponi2024autonomous, mei2024transferring}.

Two navigation tasks were implemented to evaluate system performance under varying anatomical constraints, as illustrated in Fig. \ ref {fig:Deployment}(b). Task 1 required autonomous guidewire advancement from start point S1 to goal point G1, traversing multiple bifurcations along a complex vascular trajectory. Representative frames in Fig. \ ref {fig:Deployment}(c) demonstrate the tip’s progression across key locations, where the predicted guidewire positions (green boxes) closely align with manually annotated ground truth labels (orange boxes). This task assessed the system’s real-time tracking capacity in response to frequent direction changes and complex curvature. The tracking algorithm achieved a mean localization error of 0.708 mm with a standard deviation of 0.695 mm while maintaining a processing rate of 38 frames per second.

Task 2, shown in Fig.~\ref{fig:Deployment}(d), involved a trajectory from start point S2 to goal point G2. Although the vascular path in this scenario involved fewer bifurcations, it introduced the requirement for precise retraction maneuvers in addition to forward advancement. This task evaluated the system’s ability to regulate bidirectional tip control under constrained geometries. Throughout the sequence, the predicted tip positions exhibited close spatial agreement with annotated ground truth, confirming consistent accuracy during both the insertion and retraction phases. The algorithm achieved a reduced mean tracking error of 0.148 mm with a standard deviation of 0.057 mm and sustained real-time performance at 36 frames per second.

Across both tasks, the visual comparison of predicted and annotated positions, alongside temporal frame-wise tracking analysis, confirms the system’s capacity to support autonomous guidewire navigation. The tracking output serves as a reliable control signal for downstream bifurcation-aware navigation planning, thereby validating the integration of vision-based perception within robotic-assisted vascular interventions.

\section{Discussions}
The Siamese network-based framework enhances guidewire tip tracking in endovascular procedures. Experimental results indicate a mean localization error of 0.421 ± 0.138 mm, outperforming prior methods in accuracy and robustness at 57.2 FPS. A key innovation lies in the hybrid attention module. This module dynamically integrates self- and cross-attention strategies tailored for vascular interventions. It leverages prior knowledge of vascular structures and guidewire movement patterns. The design enables the differentiation of guidewire tips from similar anatomical features in DSA images. The dual attention mechanism manages visual ambiguities and dynamic deformations effectively. This results in robust feature extraction and representation learning, achieving a mean IoU of 0.782 across diverse clinical scenarios. The low standard deviation of 0.138 mm reflects stable tracking across frames, crucial for dynamic DSA imaging.

A robotic guidewire navigation system demonstrates reliable autonomous performance, with errors of 0.708 mm ± 0.695 mm in Task 1 and 0.148 mm ± 0.057 mm in Task 2, respectively. It manages complex vascular anatomies and precise positioning at 36-38 FPS, down from 57.2 FPS in DSA due to computational demands. Variability stems from rapid guidewire motion, inconsistent contrast, and occlusions in Task 1's complex paths versus Task 2's simpler ones. Using 2D bounding box coordinates from RGB images, the system aligns with real-time clinical imaging and corrects actuation errors, enhancing its practical utility beyond static DSA tracking.

Despite these results, limitations persist. Validation relies on a proprietary dataset due to the absence of public DSA tracking datasets. It currently spans only three clinical sequences. Testing on a larger dataset would strengthen generalizability. The zero-shot deployment to the robotic RGB system encounters a domain gap. Preprocessing RGB images with strong exposure, high contrast, and grayscale enhances guidewire visibility by mimicking DSA. However, generalization remains limited without cross-modal validation or fine-tuning, given imaging disparities. The 0.148 mm error in Task 2 reflects clear, low-noise images, differing from clinical DSA, while Task 1's higher error highlights challenges in complex scenarios. Performance under extreme conditions, such as rapid guidewire motion or severe artifacts, needs further study. Enhanced vascular structure analysis and guidewire behavior modeling could improve future iterations. Additional validation in complex clinical scenarios and diverse vascular anatomies would provide a fuller evaluation. Future studies will address these with broader datasets and rigorous cross-modal validation.

The current architecture effectively leverages spatial feature extraction through template matching and attention mechanisms, but its handling of temporal dependencies is limited to frame-to-frame matching, which may not fully capture the significant motion and deformation in DSA sequences. To address this limitation, integrating an RNN-based or Transformer-based temporal attention mechanism represents a promising future direction. Such a module could process feature sequences from multiple frames to model cross-frame temporal relationships, enhancing the framework’s robustness to rapid guidewire motion and vascular deformations. Both RNN-based and Transformer-based approaches offer potential to improve temporal consistency, with RNNs capturing sequential dynamics and Transformers modeling long-range dependencies. However, challenges such as increased computational complexity, particularly for real-time applications, would need to be addressed, potentially through efficient attention mechanisms or lightweight architectures. Developing this temporal attention structure could further reduce localization errors in dynamic clinical scenarios, particularly under conditions involving severe motion or occlusions, and will be a key focus of future research to advance the framework’s performance.

While the framework achieves a processing speed of 57.2 FPS, satisfying real-time requirements, its computational complexity and resource consumption are critical considerations for deployment on robotic platforms with constrained hardware. The network’s complexity arises primarily from the Siamese ResNet-50 backbone, which dominates the computational load due to its deep convolutional layers, followed by the multi-scale feature extraction, Transformer encoder, and DEAN module, each contributing additional processing demands. These components require significant memory and computational resources, posing challenges for embedded systems commonly used in robotic platforms for endovascular procedures. The observed reduction to 36-38 FPS in robotic tasks reflects additional overhead from real-time actuation, sensor integration, and control systems. To enhance feasibility, optimization strategies such as model pruning, quantization, and efficient inference pipelines can reduce computational and memory demands while preserving real-time performance. Future work will focus on developing lightweight architectures and hardware-specific optimizations to ensure efficient deployment on robotic platforms, thereby validating the framework’s practical utility in resource-constrained clinical settings.

\section{Conclusion and Future Work}
This paper presents a novel deep-learning framework for guidewire tip tracking in image-guided endovascular procedures, demonstrating sub-pixel accuracy and robust performance across diverse clinical scenarios. By integrating dual attention mechanisms with a Siamese architecture, our approach effectively addresses the inherent challenges of endovascular procedures while maintaining real-time processing capabilities essential for clinical applications. Extensive experimental validations demonstrate superior tracking accuracy and computational efficiency that meet clinical requirements. Furthermore, successful integration with a robotic guidewire manipulation system validates the framework's practical utility in autonomous navigation tasks, showcasing its potential for advancing robot-assisted endovascular interventions.

Future research directions will focus on expanding the validation to larger clinical datasets encompassing more diverse pathological conditions and intervention types. The incorporation of respiratory and cardiac motion compensation mechanisms presents a promising avenue for enhancing tracking stability in dynamic environments. Additionally, the development of adaptive learning strategies will be crucial for handling extreme cases and unexpected scenarios during interventions. The integration of this tracking system with robotic navigation platforms and the investigation of multi-modal fusion techniques could further advance image-guided robotic intervention \cite{ wei2024absolute}. These developments aim to establish more sophisticated interventional assistance systems, potentially improving the safety and efficiency of endovascular procedures through enhanced visualization and control capabilities \cite{mei2024transferring}.

\section*{Acknowledgment}
This work is supported by the National Key Research and Development Program of China under Grant No. 2023YFB4705200, the National Natural Science Foundation of China under Grant No. 62273257, and the Open Project Fund of the State Key Laboratory of Cardiovascular Diseases under Grant No. 2024SKL-TJ002. The study protocol was reviewed and approved by the Ethics Committee of the First Affiliated Hospital of Nanjing Medical University (Approval No. 2024-SR-462), and all procedures were conducted in accordance with the Declaration of Helsinki.

\section*{Declaration of Interest}
The authors have no financial or personal conflicts of interest that could influence this work.

\bibliographystyle{ieeetr}
\balance
\bibliography{reference}

\end{document}